\documentclass[prd,preprint,tightenlines,preprintnumbers,superscriptaddress,amsmath,amssymb]{revtex4-1}
\usepackage[dvips,final]{graphicx}
\usepackage{amssymb}
\usepackage{amsmath}
\usepackage{amsfonts}
\usepackage{epsfig}
\usepackage{bm}
\usepackage{appendix}
\usepackage[section]{placeins}
\usepackage{multirow}
\usepackage{booktabs}
\usepackage{array}
\usepackage{tabularx}
\usepackage{pstricks}
\usepackage{mathrsfs}
\usepackage{cancel}
\usepackage{accents}
\usepackage{amssymb,cancel,amsmath,relsize}
\usepackage{dcolumn}
\usepackage[caption=false]{subfig} 
\usepackage{physics}
\usepackage{feynmp-auto}
\usepackage[T1]{fontenc}	
\usepackage{xcolor}
\usepackage{hyperref}
\usepackage[capitalise]{cleveref}
\usepackage{booktabs}
\usepackage[utf8]{inputenc}
\hypersetup{
    colorlinks,
    linkcolor={red!50!black},
    citecolor={blue!50!black},
    urlcolor={blue!80!black}
}
\usepackage[normalem]{ulem}
\usepackage{cleveref}
\usepackage{fancybox}

\graphicspath{{figs/}}

\begin{document}

\title{
Minimal Scale-Invariant Dark Matter}

\author{Alexander Belyaev }
\email{a.belyaev@soton.ac.uk}
\affiliation{School of Physics and Astronomy, University of Southampton, Highfield, Southampton SO17 1BJ, UK}
\affiliation{Particle Physics Department, Rutherford Appleton Laboratory, Chilton, Didcot, Oxon OX11 0QX, UK}

\author{Roman Pasechnik}
\email{Roman.Pasechnik@cern.ch}
\affiliation{Department of Physics, Lund
University, SE-223 62 Lund, Sweden}

\author{Rishav Roshan}
\email{r.roshan@soton.ac.uk}
\affiliation{School of Physics and Astronomy, University of Southampton,
Southampton, United Kingdom}

\author{Alfonso Zerwekh}
\email{alfonso.zerwekh@usm.cl}
\affiliation{Departamento de Física and Centro Cient\'ifico-Tecnol\'ogico de Valpara\'iso,  Universidad T\'ecnica Federico Santa Mar\'ia, Avenida Espa\~na 1680, Valpara\'iso, Chile}
\affiliation{Millennium Institute for Subatomic Physics at High Energy Frontier – SAPHIR,
Fernandez Concha 700, Santiago, Chile}

\begin{abstract}

We study the minimal classically scale-invariant extension of the Standard Model, containing a single $\mathbb Z_2$-stabilised real scalar singlet whose mass is not an independent input but is generated dynamically through the quantum effective potential and linked to radiative electroweak symmetry breaking through the Higgs portal. We construct the full two-field one-loop effective potential and introduce a modified on-shell renormalisation scheme that fixes the electroweak vacuum, the Higgs mass, the vanishing Higgs--singlet mixing and the singlet curvature at the physical point, yielding predictions stable under renormalisation-scale variation. Since thermal freeze-out is excluded by direct-detection limits, we identify a highly predictive freeze-in realisation of this model. Imposing perturbativity, vacuum stability and the observed relic abundance leads to a freeze-in solution with a dark-matter mass of around $2~{\rm MeV}$. The predicted electron-scattering cross section for this solution lies far below the current sensitivity of DAMIC-M. Current direct-detection experiments therefore do not constrain this scenario. The minimal scale-invariant singlet model thus provides a robust and highly predictive framework connecting radiative electroweak symmetry breaking and freeze-in dark-matter genesis.

\end{abstract}

\maketitle

\tableofcontents

\section{Introduction}

The origin of the electroweak and dark-matter mass scales remains one of the central open questions in particle physics. In the Standard Model (SM), the electroweak scale enters as an explicit dimensionful parameter through the quadratic Higgs term, and most particle dark-matter (DM) models introduce the DM mass as yet another independent input scale. Classical scale invariance offers a more restrictive framework in which explicit dimensionful parameters are absent from the scalar potential and physical scales arise radiatively through dimensional transmutation, following the Coleman--Weinberg mechanism~\cite{Coleman:1973jx,Gildener:1976ih}. In this setting, electroweak symmetry breaking (EWSB) and dark-matter mass generation originate from the same quantum effective potential.

The phenomenology of classically scale-invariant extensions of the SM has been explored extensively. Early proposals established that a minimal real singlet coupled to the Higgs can trigger radiative EWSB \cite{Foot:2007as,Foot:2007iy,Meissner:2006zh,Endo:2015ifa}, while subsequent works embedded the dark sector within hidden gauge groups or additional scalars, demonstrating that the same dimensional-transmutation dynamics can simultaneously fix the electroweak scale and the DM relic abundance \cite{Hambye:2013dgv,Gabrielli:2013hma,Demir:2013uja,Endo:2015nba}. Detailed analyses of $U(1)$ and non-Abelian hidden-sector models \cite{Karam:2015jta,Karam:2016rsz} subsequently mapped out the regions consistent with relic-density, direct-detection and collider constraints, and the role of classical scale invariance in linking dark matter to the electroweak phase transition was investigated in \cite{Plascencia:2015xwa}. In particular, Ref.~\cite{Endo:2015nba} showed that the minimal classically scale-invariant SM extension with a scalar singlet can reproduce the observed relic density, but only at the expense of a very large Higgs portal coupling, which places the DM in thermal equilibrium in the early Universe and produces it via the freeze-out mechanism \cite{Kolb:1990vq, Jungman:1995df, Bertone:2004pz, Feng:2010gw, Arcadi:2017kky, Roszkowski:2017nbc}. Such large portal couplings are excluded by the current direct-detection bounds \cite{LZ:2024zvo}. Building on this, we show for the first time that the same minimal extension can accommodate a MeV-scale DM candidate produced through the freeze-in mechanism \cite{Hall:2009bx,Bernal:2017kxu,Barman:2020plp,Barman:2021tgt,Datta:2021elq}, in which the DM never reaches thermal equilibrium with the SM plasma and is instead generated slowly from the thermal bath.

The minimal dark-matter realisation extends the Standard Model by a single real scalar singlet $S$, stabilised by an exact $\mathbb Z_2$ symmetry. Classical scale invariance forbids a bare singlet mass. Radiative symmetry breaking generates the electroweak scale through the quantum effective potential, and the Higgs portal then induces the singlet mass dynamically. At one loop, the physical singlet mass and the interaction vertices are determined by \emph{different} derivatives of the renormalised effective potential: dark-matter production and scattering are controlled by an effective portal coupling extracted from the renormalised $hSS$ vertex, while the physical mass is fixed by the curvature of the potential in the singlet direction (Sec.~\ref{subsec:effective-couplings}). This distinction is essential for connecting the quantum scalar potential consistently to dark-matter phenomenology.

The first central result of this work is the construction of a renormalised one-loop framework in which the physical predictions are stable under variations of the renormalisation scale. The one-loop effective potential contains explicit logarithmic dependence on $\mu$, which would otherwise propagate into the vacuum structure and the effective couplings. We impose a modified on-shell prescription in which the finite counterterms are fixed by physical conditions at the electroweak vacuum. The vacuum expectation value, the Higgs mass, the vanishing Higgs--singlet mixing and the singlet curvature are preserved when the renormalisation scale is varied. The remaining scale dependence affects only quantities not fixed by these conditions and represents the expected higher-order uncertainty of the one-loop truncation.

This prescription differs from the strict Gildener--Weinberg construction. In the conventional Gildener--Weinberg treatment, the mass of the scalar associated with the flat direction is generated as a one-loop prediction. In the modified on-shell formulation adopted here, the measured values of $v$ and $m_h$ are physical renormalisation inputs. The predictive power of the model instead follows from the absence of an independent singlet mass parameter and from the simultaneous requirements of quantum scale invariance, positive effective couplings, perturbative control, global vacuum stability and the observed dark-matter abundance.

The conventional thermal freeze-out realisation is excluded. In that regime, the portal interaction required to reproduce the observed relic abundance also predicts a spin-independent DM--nucleon scattering rate incompatible with current direct-detection limits~\cite{Endo:2015nba,LZ:2022lsv}. The theoretically viable regions found from the renormalised potential instead occur at extremely small positive values of $\lambda_{HS}^{\rm eff}$. Such couplings prevent the singlet from thermalising with the Standard Model plasma and place the surviving model in the freeze-in regime.

The second central result of this work is the identification of the corresponding viable and highly predictive freeze-in solution. While the theoretical requirements of $\lambda_S^{\rm eff}>0$, perturbative unitarity, and global vacuum stability allow two viable regions of parameter space, corresponding to light and heavy dark-matter masses, the freeze-in mechanism selects only the light dark-matter region. The low-mass region is determined by two nearby roots of $\lambda_S^{\rm eff}=0$, which enclose a very narrow interval where $\lambda_S^{\rm eff}>0$. The relic-density condition crosses this interval at $m_S\simeq2.59$~MeV and $2.68$~MeV, thereby fixing the light solution. The heavy surviving region opens near $m_S\simeq436$~GeV, where $\lambda_S^{\rm eff}$ turns positive, and closes near $m_S\simeq632$~GeV, where the one-dimensional vacuum-energy condition $V_{\rm eff}(v,0)=V_{\rm eff}(0,0)$ along the electroweak ray is met. One might naively expect the relic abundance to be reproducible across this entire heavy mass
range through the freeze-in mechanism with $\lambda_{HS}^{\rm eff} \sim 10^{-11}$. However,
the required tree-level portal coupling lies in the range $\lambda_{HS} \simeq 6.3$--$13.2$,
and obtaining such a small $\lambda_{HS}^{\rm eff}$ from these inputs would require
an enormous and unnatural cancellation. Therefore, the higher-order corrections to the 
portal coupling can potentially bring the singlet into thermal equilibrium,
so it would freeze out rather than freeze in. Taken together, the quantum, vacuum, and cosmological requirements therefore favour the lighter freeze-in solution over the heavier one.

For the MeV solution, the Higgs-mediated interaction can be expressed in the standard DM--electron scattering formalism. Since the momentum transfer is much smaller than the Higgs mass, the interaction is contact-like and corresponds to $F_{\rm DM}(q)=1$. For this benchmark, the predicted reference cross section, $\bar\sigma_e\simeq5.5\times10^{-65}~{\rm cm}^2$, lies approximately twenty-eight orders of magnitude below the present DAMIC-M sensitivity at $m_S\simeq2.6~{\rm MeV}$~\cite{DAMIC-M:2025luv}. Current electron-recoil experiments therefore do not constrain the light solution.

The remainder of the paper is organised as follows. Section~\ref{sec:model} develops the renormalised one-loop framework and defines the effective couplings. Section~\ref{sec:dark-matter} determines the viable freeze-in solutions and discusses the light- and heavy-mass regions. The technical details of the effective potential and counterterm construction are collected in the Appendix. We conclude in Sec.~\ref{sec:conclusions}.

\section{Scale-invariant $\mathbb Z_2$-symmetric singlet model}
\label{sec:model}
\subsection{Tree-level potential and vacuum structure}
\label{subsec:model-vacuum}

We consider the classically scale-invariant real-singlet extension of the Standard Model, supplemented by an exact $\mathbb Z_2$ symmetry under which
\begin{equation}
    S \longrightarrow -S,\qquad H \longrightarrow H .
\end{equation}
Classical scale invariance forbids explicit dimensionful parameters in the scalar potential, so that all mass scales must arise through dimensional transmutation, as in the Coleman--Weinberg mechanism~\cite{Coleman:1973jx}. The most general renormalisable scalar potential consistent with the gauge symmetry and the $\mathbb Z_2$ parity is

\begin{equation}
    V_0(H,S)
    =
    \lambda_H (H^\dagger H)^2
    + \frac{\lambda_{HS}}{2} (H^\dagger H) S^2
    + \frac{\lambda_S}{4} S^4 .
    \label{eq:V0_HS}
\end{equation}
In terms of the neutral Higgs background $\phi$ and the real singlet background $\varphi$,
\begin{equation}
    H =
    \frac{1}{\sqrt{2}}
    \begin{pmatrix}
        0 \\ \phi
    \end{pmatrix},
    \qquad
    S = \varphi,
    \label{eq:background_fields}
\end{equation}
the tree-level background potential reads
\begin{equation}
    V_0(\phi,\varphi)
    =
    \frac{\lambda_H}{4}\phi^4
    +
    \frac{\lambda_{HS}}{4}\phi^2\varphi^2
    +
    \frac{\lambda_S}{4}\varphi^4 .
    \label{eq:V0_phivarphi}
\end{equation}
We focus on the branch of the theory in which the electroweak vacuum preserves the $\mathbb Z_2$ symmetry,
\begin{equation}
    \langle \phi \rangle = v,
    \qquad
    \langle \varphi \rangle = v_s =0,
    \label{eq:vacuum_choice}
\end{equation}
with $v\simeq 246~{\rm GeV}$. In this vacuum the singlet does not mix with the neutral Higgs excitation and remains stable; it is therefore the dark-matter candidate. This is the minimal Higgs-portal dark-matter realisation in the classically scale-invariant singlet model~\cite{Endo:2015nba}. In the renormalisation prescription specified below, the position of the vacuum is kept fixed order by order, so that the $\mathbb Z_2$ symmetry is preserved by the renormalised one-loop effective potential at the physical vacuum.

Since Eq.~\eqref{eq:V0_phivarphi} is homogeneous of degree four, Euler's theorem gives
\begin{equation}
    \phi\frac{\partial V_0}{\partial \phi}
    +
    \varphi\frac{\partial V_0}{\partial \varphi}
    =
    4V_0,
    \qquad
    V_0
    =
    \frac14
    \left(
        \phi\frac{\partial V_0}{\partial \phi}
        +
        \varphi\frac{\partial V_0}{\partial \varphi}
    \right).
    \label{eq:euler_theorem}
\end{equation}
The identity applies to the full quartic potential, including the portal term. Hence any stationary ray of the tree-level scale-invariant potential with non-vanishing field value has vanishing tree-level vacuum energy. At the $\mathbb Z_2$-preserving point $(\phi,\varphi)=(v,0)$, the tree-level stationarity conditions are
\begin{align}
    \left.
    \frac{\partial V_0}{\partial \phi}
    \right|_{(v,0)}
    &=
    \lambda_H v^3,
    &
    \left.
    \frac{\partial V_0}{\partial \varphi}
    \right|_{(v,0)}
    &=0 .
    \label{eq:tree_tadpoles}
\end{align}
Thus, for $v\neq0$, the classical flat-direction condition requires
\begin{equation}
    \lambda_H(\mu_\star)=0,
    \label{eq:lambdaH_GW_scale}
\end{equation}
at a special renormalisation scale $\mu_\star$, which we identify with the Gildener--Weinberg scale~\cite{Gildener:1976ih}. Along this ray,
\begin{equation}
    V_0(\phi,0)=0
    \qquad
    \text{for}
    \qquad
    \lambda_H(\mu_\star)=0,
    \label{eq:tree_flat_direction}
\end{equation}
and the electroweak scale and the curvature along the flat direction are selected by loop effects.

With the portal normalisation in Eq.~\eqref{eq:V0_phivarphi}, the tree-level scalar curvature matrix is
\begin{equation}
    \mathcal M_0^2(\phi,\varphi)
    =
    \begin{pmatrix}
        3\lambda_H\phi^2+\tfrac12\lambda_{HS}\varphi^2
        &
        \lambda_{HS}\phi\varphi
        \\
        \lambda_{HS}\phi\varphi
        &
        \tfrac12\lambda_{HS}\phi^2+3\lambda_S\varphi^2
    \end{pmatrix}.
    \label{eq:tree_scalar_curvature_preliminary}
\end{equation}
With this normalisation, the off-diagonal entry is $\partial_\phi\partial_\varphi V_0=\lambda_{HS}\phi\varphi$. Part of the literature instead writes the portal as $\lambda_{HS}(H^\dagger H)S^2$; in that convention all our $\lambda_{HS}$ values are halved and the $hSS$ vertex doubled, which should be kept in mind in comparisons. At the vacuum and at $\mu_\star$,
\begin{equation}
    \mathcal M_0^2(v,0)
    =
    \begin{pmatrix}
        0 & 0 \\
        0 & \tfrac12\lambda_{HS}v^2
    \end{pmatrix}.
    \label{eq:tree_scalar_curvature_vacuum}
\end{equation}
The radial doublet mode is therefore massless at tree level and is identified with the pseudo-Goldstone boson of spontaneously broken scale invariance, while the singlet has
\begin{equation}
    m_{S,{\rm tree}}^2=\tfrac12\lambda_{HS}v^2.
    \label{eq:singlet_tree_mass}
\end{equation}
Local stability in the singlet direction requires $\lambda_{HS}>0$.

We follow the Gildener--Weinberg expansion, implemented for scale-invariant singlet extensions in Ref.~\cite{Endo:2015ifa}, in which
\begin{equation}
    \lambda_H=\mathcal O(\hbar),
    \qquad
    \lambda_H(\mu_\star)=0,
    \label{eq:GW_ES_counting}
\end{equation}
whereas $\lambda_{HS}$ and $\lambda_S$ are leading-order scalar couplings. Consequently, in the one-loop Coleman--Weinberg potential and in its derivatives, $\lambda_H$ is set to zero inside the field-dependent masses. This avoids double counting higher-order terms and consistently describes the observed Higgs boson as the loop-lifted scalon.

\subsection{Coleman--Weinberg potential and NLO power counting}
\label{subsec:CW-potential}

The one-loop effective potential is evaluated in the $\overline{\rm MS}$ scheme and in Landau gauge. In terms of the field-dependent squared masses $M_i^2(\phi,\varphi)$, the Coleman--Weinberg contribution is~\cite{Coleman:1973jx}
\begin{equation}
    V_{\rm CW}(\phi,\varphi)
    =
    \sum_i
    \frac{n_i}{64\pi^2}
    M_i^4(\phi,\varphi)
    \left[
        \log\frac{M_i^2(\phi,\varphi)}{\mu^2}
        -
        c_i
    \right],
    \label{eq:VCW_general}
\end{equation}
with
\begin{equation}
    c_i=
    \begin{cases}
        \dfrac32, & \text{scalars and fermions},\\[1ex]
        \dfrac56, & \text{gauge bosons}.
    \end{cases}
    \label{eq:CW_constants}
\end{equation}
The sum runs over
\begin{equation}
    i=\{W,Z,t,G,+,-\},
    \label{eq:CW_spectrum_set}
\end{equation}
with multiplicities
\begin{equation}
    n_W=6,
    \qquad
    n_Z=3,
    \qquad
    n_t=-12,
    \qquad
    n_G=3,
    \qquad
    n_+=n_-=1 .
    \label{eq:CW_multiplicities}
\end{equation}
Here $G$ denotes the three Goldstone modes and $\pm$ the two CP-even scalar eigenmodes. The scalar masses in Eq.~\eqref{eq:VCW_general} are understood as the NLO masses in Eqs.~\eqref{eq:goldstone_mass_NLO} and \eqref{eq:Fpm_NLO}, following Eq.~\eqref{eq:GW_ES_counting}. Thus, at the electroweak vacuum,
\begin{equation}
    m_G^2(v,0)=0,
    \qquad
    F_-^{\rm NLO}(v,0)=0,
    \qquad
    F_+^{\rm NLO}(v,0)=\tfrac12\lambda_{HS}v^2 .
    \label{eq:NLO_masses_vacuum_CW}
\end{equation}
The first two relations reflect the Goldstone modes and the classical flat direction, respectively. In the strict one-dimensional Gildener--Weinberg expansion, the self-loop of the scalon is of higher order because its mass is itself generated radiatively. In the two-field formulation used here, both eigenvalues are nevertheless retained in the formal expression for $V_{\rm CW}$ and in the derivative identities; the strict flat-direction limit is recovered by restricting the potential to the electroweak ray and expanding consistently.

We set
\begin{equation}
    \mu=\mu_\star,
    \label{eq:mu_choice}
\end{equation}
and, in numerical applications, identify $\mu_\star$ with the electroweak scale $v$. In the strict Gildener--Weinberg construction $\mu_\star$ would instead be determined by the running of $\lambda_H$; identifying it with $v$ is part of the definition of our modified scheme, and the sensitivity to this choice is of higher order by the argument below. The physical vacuum position and curvature conditions are then imposed by finite counterterms. This fixed-scale prescription is convenient for a two-field analysis because the same renormalised potential can be used near the physical vacuum and away from the electroweak ray.

The appearance of \(\mu_\star\) should therefore be understood as specifying
the fixed renormalisation scale at which the flat-direction condition and the
finite counterterm conditions are imposed.

A central feature of the renormalisation prescription adopted here is that the physical electroweak vacuum and the scalar curvatures are kept fixed under variations of the renormalisation scale at the order considered. When $\mu$ is changed, the Coleman--Weinberg contribution varies explicitly through the logarithms in Eq.~\eqref{eq:VCW_general}. The finite counterterms are then re-evaluated from the same tadpole and curvature conditions, Eqs.~\eqref{eq:modified_OS_tadpoles} and \eqref{eq:modified_OS_Hessian}. Consequently, the vacuum position, the Higgs mass, the vanishing Higgs--singlet mixing and the singlet curvature imposed at $(v,0)$ remain unchanged within the one-loop prescription. In this sense, the physical renormalisation point is stable against the arbitrary choice of $\mu$, while any remaining scale dependence is confined to quantities not fixed by these local conditions and is formally of higher perturbative order.

We do not perform an RG improvement of the full off-shell two-field potential. Instead, we work at fixed one-loop order and impose the physical renormalisation conditions independently at each chosen value of $\mu$. This distinction is important. The off-shell shape of the truncated effective potential retains a residual scale dependence, as expected at finite perturbative order, but the vacuum position and the physical curvatures fixed by the modified on-shell conditions do not. We restrict the analysis to regions where the Coleman--Weinberg logarithms remain perturbatively controlled, so that the residual scale variation of quantities not fixed by the renormalisation conditions provides an estimate of neglected higher-order corrections rather than an uncontrolled ambiguity. This procedure is the basis for the robustness of the dark-sector predictions discussed below.

As usual for perturbative effective-potential calculations, intermediate off-shell quantities are gauge dependent. The Nielsen identity controls this gauge dependence, and physical statements must be formulated in terms of consistently defined quantities at extrema or in a fixed perturbative scheme~\cite{Nielsen:1975fs,Andreassen:2014eha}. Throughout this work we use Landau gauge and regard the finite counterterm prescription below as part of the definition of our one-loop scheme.

\subsection{Modified on-shell renormalisation and scale stability}
\label{subsec:modified-OS-summary}

We define the renormalised one-loop potential by supplementing the Coleman--Weinberg contribution with finite counterterms fixed at the physical electroweak vacuum. The complete derivation, including the field-dependent spectrum, derivative formulae and explicit counterterm solutions, is given in Appendix~\ref{app:technical-effective-potential}. The renormalisation conditions are
\begin{equation}
    \left.
    \partial_{\chi_i}(V_{\rm CW}+V_{\rm CT})
    \right|_{(v,0)}=0,
    \qquad
    \chi_i\in\{\phi,\varphi\},
    \label{eq:modified_OS_tadpoles_main}
\end{equation}
and
\begin{equation}
    \left.
    \partial_{\chi_i}\partial_{\chi_j}(V_{\rm CW}+V_{\rm CT})
    \right|_{(v,0)}
    =
    \begin{pmatrix}
        m_h^2 & 0\\
        0 & 0
    \end{pmatrix}_{ij}.
    \label{eq:modified_OS_Hessian_main}
\end{equation}
The first condition keeps the vacuum at $(v,0)$, while the second lifts the flat Higgs direction to the measured Higgs mass, preserves the vanishing Higgs--singlet mixing and leaves the tree-level singlet curvature unchanged.

The treatment of the renormalisation scale is central to our construction. When $\mu$ is varied, the explicit logarithmic dependence of $V_{\rm CW}$ changes, and the finite counterterms are recalculated from the same physical conditions in Eqs.~\eqref{eq:modified_OS_tadpoles_main} and \eqref{eq:modified_OS_Hessian_main}. The vacuum position and the imposed scalar curvatures therefore remain fixed at the one-loop order considered. The off-shell potential retains the residual scale dependence expected from a finite-order calculation, but this dependence affects only quantities not fixed by the local renormalisation conditions and provides an estimate of neglected higher-order corrections. This prescription is the basis for the robustness of the dark-sector predictions derived below.

The construction differs from the strict Gildener--Weinberg treatment, in which the mass of the scalar along the flat direction is a one-loop prediction. Here, $v$ and $m_h$ are physical renormalisation inputs. The predictive content instead follows from the absence of an independent singlet mass and from the simultaneous imposition of the scale-invariant vacuum conditions, perturbative consistency, global vacuum stability and the observed dark-matter abundance.

\subsection{Renormalised two-field effective potential}
\label{subsec:renormalised-effective-potential}

The renormalised one-loop effective potential is
\begin{equation}
    V_{\rm eff}(\phi,\varphi)
    =
    V_0(\phi,\varphi)
    +
    V_{\rm CW}^{\rm NLO}(\phi,\varphi)
    +
    V_{\rm CT}(\phi,\varphi),
    \label{eq:Veff_full_definition}
\end{equation}
where $V_0$ is given in Eq.~\eqref{eq:V0_phivarphi}, $V_{\rm CW}^{\rm NLO}$ in Sec.~\ref{subsec:CW-potential}, and $V_{\rm CT}$ in Eq.~\eqref{eq:VCT_final_minimal}. By construction,
\begin{equation}
    \left.\partial_\phi V_{\rm eff}\right|_{(v,0)}=0,
    \qquad
    \left.\partial_\varphi V_{\rm eff}\right|_{(v,0)}=0,
    \label{eq:Veff_stationarity}
\end{equation}
and
\begin{equation}
    \left.
    \begin{pmatrix}
        \partial_\phi^2V_{\rm eff}
        &
        \partial_\phi\partial_\varphi V_{\rm eff}
        \\
        \partial_\varphi\partial_\phi V_{\rm eff}
        &
        \partial_\varphi^2V_{\rm eff}
    \end{pmatrix}
    \right|_{(v,0)}
    =
    \begin{pmatrix}
        m_h^2 & 0\\
        0 & m_S^2
    \end{pmatrix},
    \qquad
    m_S^2=\tfrac12\lambda_{HS}v^2.
    \label{eq:Veff_Hessian_vacuum}
\end{equation}
The vanishing off-diagonal entry follows from the $\mathbb Z_2$-preserving vacuum, while the second diagonal entry follows from Eq.~\eqref{eq:singlet_curvature_condition}. The Goldstone curvature is also preserved,
\begin{equation}
    \left.
    \frac{\partial^2V_{\rm eff}}{\partial G^0\partial G^0}
    \right|_{(v,0)}=0,
    \label{eq:Veff_Goldstone_curvature}
\end{equation}
and similarly for the charged Goldstone modes.

Equations~\eqref{eq:Veff_stationarity}--\eqref{eq:Veff_Goldstone_curvature} are local renormalisation conditions. They do not by themselves guarantee that $(v,0)$ is the global minimum of the one-loop potential. The global selection of the $\mathbb Z_2$-preserving electroweak vacuum is imposed in Sec.~\ref{subsec:boundedness-vacuum-selection}. Since the potential is truncated at one loop, finite counterterm choices that are equivalent at the imposed renormalisation point differ away from the vacuum by higher-order terms. All numerical results are therefore quoted in the finite scheme specified by Eq.~\eqref{eq:minimal_singlet_counterterm_convention}.

\subsection{Boundedness, perturbativity and vacuum selection}
\label{subsec:boundedness-vacuum-selection}

The local conditions above must be supplemented by boundedness, perturbativity and global vacuum-selection requirements. At tree level, the quartic potential in Eq.~\eqref{eq:V0_phivarphi} is bounded from below if the corresponding copositivity conditions are satisfied~\cite{Kannike:2012pe}:
\begin{equation}
    \lambda_H\geq0,
    \qquad
    \lambda_S\geq0,
    \qquad
    \tfrac12\lambda_{HS}+\sqrt{\lambda_H\lambda_S}\geq0 .
    \label{eq:tree_boundedness_general}
\end{equation}
At the Gildener--Weinberg scale, where $\lambda_H(\mu_\star)=0$, these reduce to
\begin{equation}
    \lambda_S\geq0,
    \qquad
    \lambda_{HS}\geq0.
    \label{eq:tree_boundedness_GW}
\end{equation}
For the phenomenologically relevant dark-matter branch we require the strict conditions
\begin{equation}
    \lambda_S>0,
    \qquad
    \lambda_{HS}>0,
    \label{eq:strict_tree_boundedness_GW}
\end{equation}
so that the singlet direction is stabilised and $m_S^2=\tfrac12\lambda_{HS}v^2>0$.

The tree-level conditions are necessary but not sufficient once the Coleman--Weinberg contribution and finite counterterms are included. We therefore impose the global vacuum requirement on the renormalised two-field potential by demanding that
\((v,0)\) is deeper than any competing stationary point of the
renormalised two-field potential,
\begin{equation}
    V_{\rm eff}(v,0)
    <
    V_{\rm eff}(\phi_{\rm st},\varphi_{\rm st}) \,,
    \label{eq:global_minimum_all_stationary}
\end{equation}
for every competing stationary point $(\phi_{\rm st},\varphi_{\rm st})\neq(v,0)$ within the perturbatively reliable field domain. In particular, any deeper extremum with $\varphi_{\rm st}\neq0$ would spontaneously break the $\mathbb Z_2$ symmetry and invalidate the singlet dark-matter interpretation. Along the electroweak ray, Eq.~\eqref{eq:analytic_vacuum_depth_condition} provides a simple analytic check of electroweak symmetry breaking, but the numerical analysis uses the full two-field condition in Eq.~\eqref{eq:global_minimum_all_stationary}.

Perturbative control further requires the scalar quartic couplings and the finite counterterm shifts to remain in the perturbative regime,
\begin{equation}
    |\lambda_{HS}|,
    \ |\lambda_S|,
    \ |\delta\lambda_H|,
    \ |\delta\lambda_{HS}|,
    \ |\delta\lambda_S|
    \ll 4\pi.
    \label{eq:perturbativity_generic}
\end{equation}
In numerical scans this is implemented through a conservative upper bound, for example
\begin{equation}
    |\lambda_i|<\lambda_{\rm max},
    \qquad
    \lambda_{\rm max}=4\pi
    \quad\text{or}\quad
    2\pi,
    \label{eq:perturbativity_numeric}
\end{equation}
depending on the desired degree of perturbative control. Points for which the one-loop correction becomes comparable to, or larger than, the leading contribution in the region relevant for vacuum selection are discarded.

Finally, where field-dependent squared masses become negative, the logarithms in $V_{\rm CW}$ develop imaginary parts. These imaginary parts signal the instability of the chosen background rather than a correction to the static potential. In the vacuum-selection analysis we use
\begin{equation}
    V_{\rm eff}\equiv {\rm Re}\,V_{\rm eff},
    \label{eq:real_part_Veff}
\end{equation}
and require the physical vacuum itself to have the non-negative curvatures
\begin{equation}
    m_h^2>0,
    \qquad
    m_S^2=\tfrac12\lambda_{HS}v^2>0,
    \qquad
    m_G^2(v,0)=0.
    \label{eq:physical_curvature_conditions}
\end{equation}
These requirements define the perturbatively reliable parameter space used in the phenomenological sections.

\begin{figure}[htb!]
\centering
\includegraphics[width=0.62\textwidth]{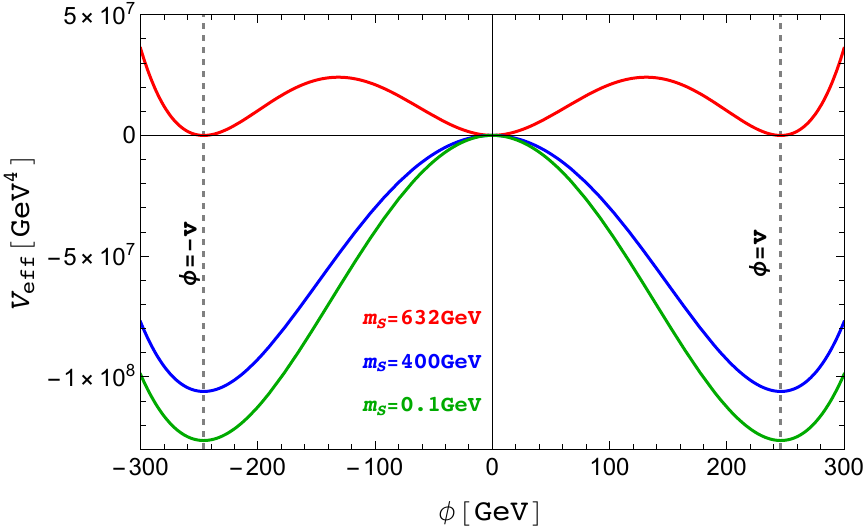}
\caption{The renormalised effective potential $V_{\rm eff}(\phi,0)$ as a function of $\phi$ for three values of the singlet mass, $m_S=632$, $400$ and $0.1$~GeV. The grey dashed lines mark $\phi=\pm v$; the red curve is close to the one-dimensional vacuum-degeneracy point. With the mass normalisation $m_S^2=\lambda_{HS}v^2/2$, the analytic condition $V_{\rm eff}(v,0)=V_{\rm eff}(0,0)$ gives $m_S\simeq632$~GeV for the numerical inputs used in the LanHEP/CalcHEP implementation, corresponding to $\lambda_{HS}\simeq13.2$. }
\label{fig:effective_potential}
\end{figure}

In Fig.~\ref{fig:effective_potential} we show the renormalised effective potential $V_{\rm eff}(\phi,0)$ along the electroweak ray. The grey dashed vertical lines mark $\phi=\pm v$, and the renormalisation scale is fixed at $\mu_\star=v$. With the portal normalisation of Eq.~\eqref{eq:V0_HS}, the one-dimensional analytic condition $V_{\rm eff}(v,0)=V_{\rm eff}(0,0)$ is reached at $m_S\simeq632$~GeV, corresponding to $\lambda_{HS}\simeq13.2$.

\subsection{Effective couplings}
\label{subsec:effective-couplings}
The dark-matter analysis below is controlled by effective couplings defined from derivatives of the renormalised potential at the physical vacuum,
\begin{equation}
\lambda_{HS}^{\rm eff}
\;\equiv\;
\frac{1}{v}
\left.
\frac{\partial^{3}V_{\rm eff}}
{\partial\phi\,\partial\varphi^{2}}
\right|_{(v,0)},
\qquad
\lambda_{S}^{\rm eff}
\;\equiv\;
\frac{1}{6}
\left.
\frac{\partial^{4}V_{\rm eff}}
{\partial\varphi^{4}}
\right|_{(v,0)},
\label{eq:effective-couplings-def}
\end{equation}
which reduce to $\lambda_{HS}$ and $\lambda_{S}$ at tree level. The trilinear coupling $\lambda_{HS}^{\rm eff}$ fixes the renormalised $hSS$ vertex,
${\cal L}\supset-\tfrac12\lambda_{HS}^{\rm eff}\,v\,hS^{2}$,
and therefore governs both freeze-in production via $h\to SS$ and Higgs-mediated scattering of $S$ on Standard-Model fermions. Beyond tree level it is not fixed simply by $2m_S^{2}/v^{2}$: the Coleman--Weinberg contribution and the finite counterterms modify the $hSS$ vertex, while the singlet curvature is held at its tree-level value by Eq.~\eqref{eq:modified_OS_Hessian_main}. The corresponding quartic $hhSS$ coupling, $\tfrac12\,\partial^{4}V_{\rm eff}/\partial\phi^{2}\partial\varphi^{2}\big|_{(v,0)}$, coincides with $\lambda_{HS}^{\rm eff}$ at tree level but differs from it at one loop; it plays no role in the processes considered in this work. Finally, $\lambda_{S}^{\rm eff}$ controls the one-loop stability of the singlet direction and delimits the theoretically allowed regions used below.

Some details of the LanHEP/CalcHEP implementation are given in Appendix~\ref{app:lanhep-calchep}. 
It uses the same notation as Eq.~\eqref{eq:V0_HS}.
The model implementation used in this work is publicly available in the High Energy Physics Model DataBase (HEPMDB) at
\url{https://hepmdb.soton.ac.uk/hepmdb:0826.0358}.

\section{Dark matter}
\label{sec:dark-matter}

The dark-matter analysis is based on the renormalised quantities defined in Sec.~\ref{subsec:effective-couplings}. The physical singlet mass is determined by the curvature of the renormalised effective potential, while production and scattering amplitudes are controlled by the effective Higgs-portal coupling $\lambda_{HS}^{\rm eff}$ of Eq.~\eqref{eq:effective-couplings-def}.

We first determine the theoretically viable parameter space. We require
\begin{equation}
\lambda_S^{\rm eff}>0,
\end{equation}
perturbative couplings and counterterms, and a physical electroweak vacuum satisfying
\begin{equation}
V_{\rm eff}(v,0)<V_{\rm eff}(0,0)=0
\end{equation}
and lying below any competing stationary point. The boundaries defined by $\lambda_S^{\rm eff}=0$ are shown in blue in Fig.~\ref{fig:scalar_masses}. The relic-density condition is then imposed on the theoretically allowed regions.
The scan proceeds as follows: the independent inputs are the couplings $(\lambda_{HS},\lambda_S)$ at $\mu_\star=v$; for each point the renormalised potential determines the singlet mass, $m_S^2=\tfrac12\lambda_{HS}v^2$, and the effective couplings of Eq.~\eqref{eq:effective-couplings-def}, in terms of which the boundaries $\lambda_S^{\rm eff}=0$ and the relic-density contour are displayed in the $(m_S,\lambda_{HS}^{\rm eff})$ plane.

\subsection{Freeze-in relic density}
\label{subsec:relic-density}

We assume a negligible initial singlet abundance and require that $S$ never reaches thermal equilibrium with the SM plasma. Its number density satisfies
\begin{equation}
\dot n_S+3Hn_S=C_{\rm prod},
\end{equation}
where $C_{\rm prod}$ contains the Higgs-portal production processes. In terms of the yield $Y_S=n_S/s$ and $x=m_S/T$,
\begin{equation}
\frac{dY_S}{dx}
=
\frac{C_{\rm prod}}{sHx}.
\label{eq:freezein_boltzmann}
\end{equation}

For $m_S<m_h/2$, production is dominated by the decay $h\to SS$, with
\begin{equation}
\Gamma(h\to SS)
=
\frac{(\lambda_{HS}^{\rm eff})^2v^2}{32\pi m_h}
\sqrt{1-\frac{4m_S^2}{m_h^2}}.
\label{eq:h_to_SS_width}
\end{equation}
For $m_S>m_h/2$, this decay is closed and the abundance is generated by $2\to2$ processes involving Higgs bosons, electroweak gauge bosons and SM fermions.

\begin{figure}[htb!]
\centering
\includegraphics[width=0.62\textwidth]{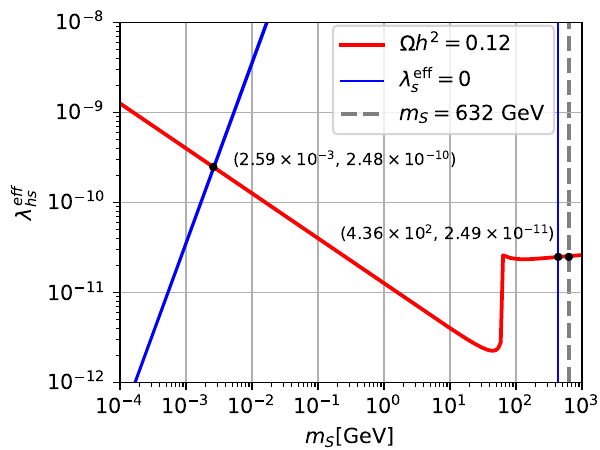}
\caption{Relic-density condition $\Omega_Sh^2=0.12$ (red) and the boundaries $\lambda_S^{\rm eff}=0$ (blue) in the $(m_S,\lambda_{HS}^{\rm eff})$ plane. The physical regions satisfy $\lambda_S^{\rm eff}>0$, perturbativity and the global-vacuum condition. The light- and heavy-mass regions are shown in detail in Figs.~\ref{fig:scalar_masses_zoom_light} and \ref{fig:scalar_masses_zoom_heavy} (only one freeze-in solution survives). }
\label{fig:scalar_masses}
\end{figure}

We compute the freeze-in abundance with {\tt micrOMEGAs 7} \cite{Belanger:2026asz}, including the relevant decay and scattering channels and the quantum-statistical distributions of the bath particles~\cite{Belanger:2018mqt}. The observed abundance is imposed through
\begin{equation}
\Omega_Sh^2=0.12,
\label{eq:observed_relic}
\end{equation}
consistent with the cosmological DM density~\cite{Planck:2018vyg}. The red curve in Fig.~\ref{fig:scalar_masses} gives the value of $\lambda_{HS}^{\rm eff}$ required to reproduce this abundance.

\begin{figure}[htb!]
\centering
\includegraphics[width=0.62\textwidth]{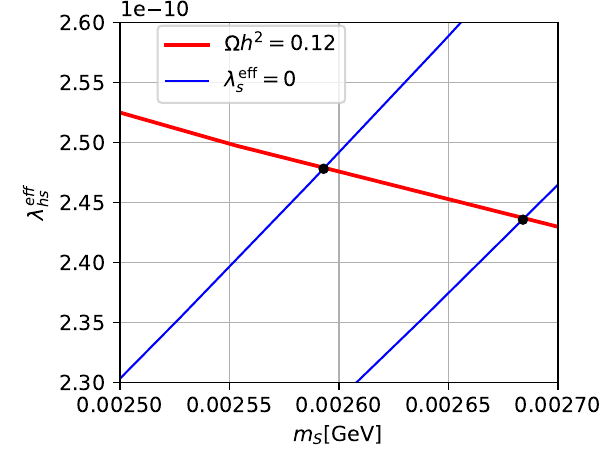}
\caption{Light-mass region. The two blue curves enclose the narrow interval where $\lambda_S^{\rm eff}>0$. Its intersections with the relic-density curve occur at $m_S\simeq2.59~{\rm MeV}$ (with $\lambda_{HS}^{\rm eff}\simeq2.48\times10^{-10}$) and $m_S\simeq2.68~{\rm MeV}$ (with $\lambda_{HS}^{\rm eff}\simeq2.43\times10^{-10}$). The two crossings are denoted by black dots. }
\label{fig:scalar_masses_zoom_light}
\end{figure}

The low-mass region is shown in Fig.~\ref{fig:scalar_masses_zoom_light}. Two nearby roots of
\begin{equation}
\lambda_S^{\rm eff}=0
\end{equation}
enclose a very narrow interval in which $\lambda_S^{\rm eff}>0$. The relic-density curve crosses this interval at the two boundary points shown by black dots in Fig.~\ref{fig:scalar_masses_zoom_light}:
 \begin{equation}
    (m_S,\lambda_{HS}^{\rm eff})_{\rm light}
\simeq (2.59~{\rm MeV},\,2.48\times10^{-10})
\quad\text{and}\quad
(2.68~{\rm MeV},\,2.43\times10^{-10}).
\label{eq:light_freezein_solution}
\end{equation}
The narrow positive-$\lambda_S^{\rm eff}$ interval and the relic-density condition therefore determine the light solution.

\begin{figure}[htb!]
\centering
\includegraphics[width=0.62\textwidth]{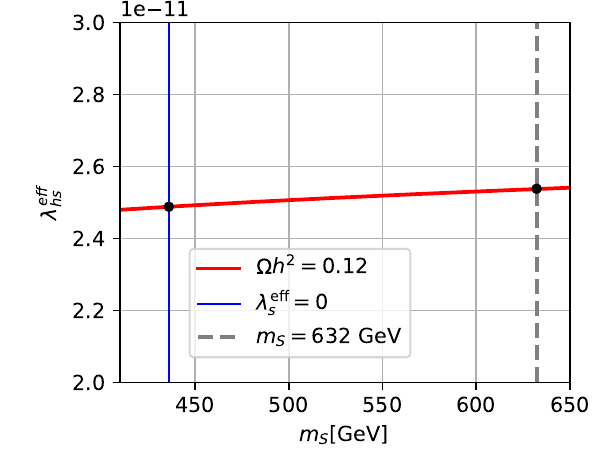}
\caption{Heavy-mass region. The blue line marks the root $\lambda_S^{\rm eff}=0$; between it and the grey dashed line, $\lambda_S^{\rm eff}>0$. The relic-density curve crosses the lower boundary near $m_S\simeq436~{\rm GeV}$.  The upper boundary, denoted by the grey dashed line, corresponds to the vacuum-degeneracy point where the one-dimensional condition $V_{\rm eff}(v,0)=V_{\rm eff}(0,0)$ is satisfied. This occurs at $m_S\simeq632~{\rm GeV}$.}   
\label{fig:scalar_masses_zoom_heavy}
\end{figure}

The high-mass region is shown in Fig.~\ref{fig:scalar_masses_zoom_heavy}.  The interval with positive singlet self-coupling begins near
\begin{equation}
m_S\simeq436~{\rm GeV},
\end{equation}
where $\lambda_S^{\rm eff}$ becomes positive, and the heavy mass regime ends near
\begin{equation}
m_S\simeq632~{\rm GeV},
\end{equation}
where the one-dimensional vacuum-energy condition
\begin{equation}
V_{\rm eff}(v,0)=V_{\rm eff}(0,0)
\end{equation}
is reached. It is interesting to point out that the observed dark-matter relic abundance could naively be reproduced throughout this heavy-mass range via the freeze-in mechanism with the effective portal coupling $\lambda_{HS}^{\rm eff}\sim \mathcal{O}(10^{-11})$ as shown by the red line in Fig.~\ref{fig:scalar_masses_zoom_heavy} (two black dots mark the boundaries). Such a small $\lambda_{HS}^{\rm eff}$ could only arise from an extremely fine-tuned cancellation, since at the input level $\lambda_{HS}$ must lie between $6.3$ and $13.2$, while the one-loop corrections shift the coupling only by $\mathcal{O}\!\left(3\lambda_{HS}(3\lambda_S+2\lambda_{HS})/16\pi^2\right)$. Therefore, portal couplings of this magnitude can potentially bring the singlet into thermal equilibrium, so it would freeze out rather than freeze in. Note also that, near the upper edge, the required couplings leave the perturbative regime of Eq. (31), $\lambda_{HS} \simeq 13.2 \gtrsim 4\pi$ with $\lambda_S^{\rm eff} \simeq 24$, so the vacuum-degeneracy bound $m_S \simeq 632$ GeV marks the limit of perturbative control. This further disfavours the heavy region, which is already incompatible with the freeze-in production solution. Moreover, along the relic-density contour the required singlet self-coupling, $\lambda_S$, is also negative throughout the entire interval $436~{\rm GeV} < m_S < 632~{\rm GeV}$. Since this violates the theoretical requirement of Eq.~\eqref{eq:strict_tree_boundedness_GW}, the heavy-mass region is excluded. The tree-level coupling $\lambda_S$ is positive
in the low-mass window where the light solution lies.  As a result, a concrete freeze-in solution of the scale-invariant model is obtained only in the narrow mass window $m_S\simeq2.59$--$2.68$~MeV, with $\lambda_{HS}^{\rm eff}\sim2.43\times10^{-10}$--$2.48\times10^{-10}$, far below the coupling required for thermal equilibrium.  For larger portal couplings, the singlet thermalises and the model enters the conventional freeze-out regime, which is excluded by DM--nucleus direct-detection limits.

The absence of thermalisation is verified by requiring
\begin{equation}
\Gamma_{S\leftrightarrow{\rm SM}}(T)<H(T)
\label{eq:no_thermalisation}
\end{equation}
over the temperature range relevant for production, where $\Gamma_{S\leftrightarrow{\rm SM}}$ includes the $h\leftrightarrow SS$ decay/inverse-decay rate and the $2\to2$ production and annihilation rates involving particles in the thermal bath.

\subsection{Direct detection}
\label{subsec:detection}

While the model offers an interesting opportunity to probe the light dark-matter solution in future direct-detection experiments such as DAMIC-M~\cite{DAMIC-M:2025luv}, we consider a representative benchmark near the centre of the allowed window:
\begin{equation}
m_S\simeq2.6~{\rm MeV},
\qquad
\lambda_{HS}^{\rm eff}\simeq2.46\times10^{-10},
\label{eq:light_DD_benchmark}
\end{equation}
For this benchmark, nuclear recoils are too small to be observed. Scattering on electrons is possible through
\begin{equation}
S e\longrightarrow S e
\end{equation}
with $t$-channel Higgs exchange.

Since the momentum transfer is much smaller than the Higgs mass, the interaction is contact-like and corresponds to~\cite{Essig:2011nj,Essig:2015cda}
\begin{equation}
F_{\rm DM}(q)=1.
\end{equation}
The appropriate comparison is therefore with the heavy-mediator limit on the reference DM--electron cross section. 

With the portal normalisation of Eq.~\eqref{eq:V0_HS}, and replacing the tree-level portal by the effective coupling, electroweak symmetry breaking gives
\begin{equation}
{\cal L}
\supset
-\frac{1}{2}\lambda_{HS}^{\rm eff}v hS^2
-\frac{m_e}{v}h\bar e e.
\label{eq:hSS_hee_interactions}
\end{equation}
The non-relativistic reference cross section is
\begin{equation}
\bar\sigma_e
=
\frac{(\lambda_{HS}^{\rm eff})^2m_e^2\mu_{Se}^2}
{4\pi m_S^2m_h^4}
=
\frac{(\lambda_{HS}^{\rm eff})^2m_e^4}
{4\pi(m_S+m_e)^2m_h^4},
\label{eq:sigmae_Higgs_portal}
\end{equation}
where
\begin{equation}
\mu_{Se}
=
\frac{m_Sm_e}{m_S+m_e}.
\end{equation}

For this benchmark,
\begin{equation}
\mu_{Se}\simeq0.427~{\rm MeV},
\end{equation}
and
\begin{equation}
\bar\sigma_e
\simeq
5.5\times10^{-65}~{\rm cm}^2.
\label{eq:sigmae_light_benchmark}
\end{equation}

The relevant experimental comparison is the heavy-mediator $F_{\rm DM}=1$ limit shown in Fig.~2 (left) of the DAMIC-M analysis~\cite{DAMIC-M:2025luv}. For the dark-matter mass range $2.59$--$2.68$~MeV, the current sensitivity is approximately
\begin{equation}
\bar\sigma_e^{\rm lim}
\sim 8.3 \times 10^{-37}~{\rm cm}^2-3.5\times 10^{-37}~{\rm cm}^2,
\end{equation}
at 90\%~C.L.
The predicted Higgs-mediated cross section is therefore approximately twenty-eight orders of magnitude below the present DAMIC-M sensitivity.

The absorption limits shown in Fig.~3 (left) of Ref.~\cite{DAMIC-M:2025luv} do not apply to this model. They constrain eV-scale hidden-photon dark matter through the kinetic-mixing parameter $\epsilon$, rather than MeV-scale scalar dark matter interacting through the Higgs portal.

The MeV-scale solution can therefore be expressed in the standard electron-recoil formalism used by DAMIC-M, SENSEI and SuperCDMS, but its predicted scattering rate is far below present experimental sensitivity. Current direct-detection experiments do not constrain this benchmark.
Three further remarks are in order. First, although the MeV-scale singlet is produced relativistically at $T\sim m_h$, its momentum redshifts by many orders of magnitude before structure formation; Lyman-$\alpha$ constraints on freeze-in from decays require $m_{\rm DM}\gtrsim{\cal O}(10)$~keV~\cite{Decant:2021mhj}, safely below our benchmark. Second, since $S$ never thermalises and its abundance only saturates the observed $\Omega_Sh^2=0.12$, its contribution to the relativistic energy density at BBN is negligible. Third, the induced invisible Higgs branching ratio is ${\rm BR}(h\to SS)\sim10^{-16}$, far below any collider sensitivity.

\section{Conclusions}
\label{sec:conclusions}

We have studied the minimal classically scale-invariant extension of the Standard Model with a single real $\mathbb Z_2$-odd scalar singlet. The model contains no explicit dark-sector mass term. Both the electroweak scale and the singlet mass are generated dynamically through quantum corrections, with the Higgs portal transmitting radiative electroweak symmetry breaking to the dark sector.

The first central result is the construction of a consistent renormalised one-loop framework for the scalar sector. We use the full two-field effective potential and impose a modified on-shell prescription in which the electroweak vacuum, the Higgs mass, the vanishing Higgs--singlet mixing and the singlet curvature remain fixed when the renormalisation scale is varied. The residual scale dependence then represents the expected higher-order uncertainty of the one-loop calculation.

Within this framework, the physical singlet mass and the effective interaction couplings are defined by different derivatives of the renormalised potential. In particular, $\lambda_{HS}^{\rm eff}$ is extracted from the renormalised $hSS$ vertex and controls dark-matter production and scattering, while $\lambda_S^{\rm eff}$ determines the stability of the singlet direction. This provides the necessary connection between the quantum scalar potential and the dark-matter analysis.

The theoretical requirements of $\lambda_S^{\rm eff}>0$, perturbative unitarity, and global vacuum stability restrict the parameter space to two regions. The first lies in the low-mass region around $m_S\simeq 2$~MeV. This freeze-in solution reproduces 
the observed relic abundance, extending along the relic-density contour between
\begin{equation}
    (m_S,\lambda_{HS}^{\rm eff})_{\rm light}
\simeq (2.59~{\rm MeV},\,2.48\times10^{-10})
\quad\text{and}\quad
(2.68~{\rm MeV},\,2.43\times10^{-10}).
\end{equation}
 
The second is the high-mass region, with $m_S$ between $436~{\rm GeV}$ and the upper bound of $632~{\rm GeV}$ imposed by the $V_{\rm eff}(v,0)=V_{\rm eff}(0,0)$ condition.
This solution, with $\lambda_{HS}^{\rm eff}\sim10^{-11}$, is theoretically disfavoured. Indeed, it requires the input Higgs-portal coupling to lie in the range
\begin{equation}
6.3\lesssim\lambda_{HS}\lesssim13.2,
\end{equation}
which demands an extreme cancellation to obtain the tiny effective coupling $\lambda_{HS}^{\rm eff}\sim10^{-11}$. Therefore, higher-loop corrections could potentially bring the singlet into thermal equilibrium.
Thus, this solution could lead to freeze-out rather than freeze-in, and determining its exclusion requires higher-order corrections that are beyond the scope of this paper.

For the MeV benchmark, Higgs exchange gives a contact-like DM--electron interaction with $F_{\rm DM}(q)=1$.  With the portal normalisation of Eq.~\eqref{eq:V0_HS} we obtain
\begin{equation}
\bar\sigma_e\simeq5.5\times10^{-65}~{\rm cm}^2.
\end{equation}
This is approximately twenty-eight orders of magnitude below the present DAMIC-M heavy-mediator sensitivity. Current DAMIC-M, SENSEI, and SuperCDMS electron-recoil searches therefore do not constrain the light solution. The absorption limits on hidden-photon dark matter are not applicable to this scalar Higgs-portal benchmark.

In summary, the minimal scale-invariant singlet model is excluded as a conventional thermal WIMP but remains viable in the freeze-in regime. The renormalised one-loop potential, the vacuum requirements and the observed relic abundance together reduce the allowed parameter space to one isolated solution. This makes the model highly predictive despite the fact that the corresponding direct-detection rates lie far below present experimental sensitivity. Intriguingly, the persistent null results of dark-matter direct-detection searches find
a natural explanation within the scenario identified here: if the minimal conformal dark
matter is realised through freeze-in, its coupling to the visible sector is necessarily so feeble
that the predicted signals lie many orders of magnitude below current and foreseeable sensitivities---the absence of a signal is a structural prediction of the mechanism rather than
an accident of parameter choice.

\section*{Acknowledgments} This work was partially funded by ANID-Chile FONDECYT
grant 1230110 and by the ANID--Chile Millennium Science Initiative Program
ICN2019\_044. A.B. and R.R. acknowledge financial support from the STFC
Consolidated Grant ST/X000583/1. A.B. is supported in part through the NExT
Institute. A.B. acknowledges partial support from Leverhulme Trust project
MONDMag (RPG-2022-57). We acknowledge the use of the IRIDIS High Performance
Computing Facility and associated support services at the University of
Southampton in the completion of this work.

\newpage
\appendix

\section{Technical details of the one-loop effective potential}
\label{app:technical-effective-potential}

This appendix collects the technical ingredients used in Sec.~\ref{subsec:modified-OS-summary}: the complete field-dependent scalar spectrum, the derivatives of the Coleman--Weinberg potential, the explicit solution of the finite counterterm conditions and the analytic flat-direction check.

\subsection{Field-dependent spectrum}
\label{subsec:field-dependent-spectrum}

The one-loop effective potential is constructed from the spectrum of field-dependent masses on a general two-field background $(\phi,\varphi)$. We work in Landau gauge and retain the electroweak gauge bosons, the top quark, the would-be Goldstone modes and the two CP-even scalar eigenmodes. Lighter SM fermions are neglected because their Yukawa couplings give numerically negligible contributions.

The gauge-boson and top-quark masses are
\begin{equation}
    m_W^2(\phi)=\frac{g^2}{4}\phi^2,
    \qquad
    m_Z^2(\phi)=\frac{g^2+g'^2}{4}\phi^2,
    \qquad
    m_t^2(\phi)=\frac{y_t^2}{2}\phi^2,
    \label{eq:field_masses_gauge_top}
\end{equation}
with multiplicities
\begin{equation}
    n_W=6,
    \qquad
    n_Z=3,
    \qquad
    n_t=-12 .
    \label{eq:multiplicities_gauge_top}
\end{equation}
The minus sign in $n_t$ accounts for fermionic statistics. The three would-be Goldstone modes have the common mass
\begin{equation}
    m_G^2(\phi,\varphi)=\lambda_H\phi^2+\tfrac12\lambda_{HS}\varphi^2,
    \qquad
    n_G=3 .
    \label{eq:goldstone_mass}
\end{equation}
They are massless on the electroweak flat direction after imposing Eq.~\eqref{eq:GW_ES_counting}.

The CP-even scalar eigenvalues are those of Eq.~\eqref{eq:tree_scalar_curvature_preliminary}. Defining
\begin{equation}
    A=3\lambda_H\phi^2+\tfrac12\lambda_{HS}\varphi^2,
    \qquad
    D=\tfrac12\lambda_{HS}\phi^2+3\lambda_S\varphi^2,
    \qquad
    B=\lambda_{HS}\phi\varphi,
    \label{eq:ADB_definitions}
\end{equation}
one obtains
\begin{equation}
    F_\pm(\phi,\varphi)
    =
    \frac12(A+D)
    \pm
    \frac12\sqrt{(A-D)^2+4B^2} .
    \label{eq:Fpm_compact}
\end{equation}
Explicitly,
\begin{align}
    F_\pm(\phi,\varphi)
    &=
    \frac12
    \left[
        (3\lambda_H+\tfrac12\lambda_{HS})\phi^2
        +
        (\tfrac12\lambda_{HS}+3\lambda_S)\varphi^2
    \right]
    \nonumber\\
    &\quad
    \pm
    \frac12
    \sqrt{
        \left[
            (3\lambda_H-\tfrac12\lambda_{HS})\phi^2
            +
            (\tfrac12\lambda_{HS}-3\lambda_S)\varphi^2
        \right]^2
        +
        4\lambda_{HS}^2\phi^2\varphi^2
    },
    \label{eq:Fpm_explicit}
\end{align}
with $n_+=n_-=1$. At $(v,0)$ and $\mu_\star$,
\begin{equation}
    F_+(v,0)=\tfrac12\lambda_{HS}v^2,
    \qquad
    F_-(v,0)=0,
    \label{eq:Fpm_vacuum}
\end{equation}
for $\lambda_{HS}>0$. Thus $F_+$ corresponds to the singlet-like scalar, whereas $F_-$ is the classically massless radial mode.

The field-dependent CP-even mixing angle may be defined by the orthogonal rotation that diagonalises Eq.~\eqref{eq:tree_scalar_curvature_preliminary}:
\begin{equation}
    \tan 2\alpha(\phi,\varphi)
    =
    \frac{2B}{D-A}
    =
    \frac{4\lambda_{HS}\phi\varphi}
    {(\lambda_{HS}-6\lambda_H)\phi^2+(6\lambda_S-\lambda_{HS})\varphi^2}.
    \label{eq:mixing_angle_background}
\end{equation}
At the $\mathbb Z_2$-preserving vacuum, $\alpha(v,0)=0$.

With the NLO prescription of Eq.~\eqref{eq:GW_ES_counting}, the scalar masses inserted into the one-loop potential are obtained by setting $\lambda_H\to0$ in Eqs.~\eqref{eq:goldstone_mass} and \eqref{eq:Fpm_explicit}:
\begin{equation}
    m_G^2(\phi,\varphi)=\tfrac12\lambda_{HS}\varphi^2,
    \label{eq:goldstone_mass_NLO}
\end{equation}
\begin{align}
    F_\pm^{\rm NLO}(\phi,\varphi)
    &=
    \frac12
    \left[
        \tfrac12\lambda_{HS}\phi^2
        +
        (\tfrac12\lambda_{HS}+3\lambda_S)\varphi^2
    \right]
    \nonumber\\
    &\quad
    \pm
    \frac12
    \sqrt{
        \left[
            -\tfrac12\lambda_{HS}\phi^2
            +
            (\tfrac12\lambda_{HS}-3\lambda_S)\varphi^2
        \right]^2
        +
        4\lambda_{HS}^2\phi^2\varphi^2
    } .
    \label{eq:Fpm_NLO}
\end{align}
This retains the full two-field dependence needed for the vacuum analysis away from the electroweak ray.

\subsection{Modified on-shell scheme}
\label{subsec:modified-OS-counterterms}

We now specify the finite counterterm prescription used to define the renormalised two-field effective potential. The counterterms impose physical renormalisation conditions adapted to the tree-level flat direction: the electroweak vacuum is kept fixed, the loop-plus-counterterm sector generates the Higgs/scalon curvature, and the singlet curvature is not shifted away from its tree-level value.

For later use, let
\begin{equation}
    x_i(\phi,\varphi)\equiv M_i^2(\phi,\varphi).
\end{equation}
Differentiating Eq.~\eqref{eq:VCW_general} gives
\begin{align}
    \partial_a V_{\rm CW}
    &=
    \sum_i
    \frac{n_i}{64\pi^2}
    2x_i\,\partial_a x_i
    \left[
        \log\frac{x_i}{\mu^2}-c_i+\frac12
    \right],
    \label{eq:VCW_first_derivative}
    \\
    \partial_a\partial_b V_{\rm CW}
    &=
    \sum_i
    \frac{n_i}{64\pi^2}
    \bigg\{
    2(\partial_a x_i)(\partial_b x_i)
    \left[
        \log\frac{x_i}{\mu^2}-c_i+\frac32
    \right]
    \nonumber\\
    &\hspace{2.6cm}
    +
    2x_i\,\partial_a\partial_b x_i
    \left[
        \log\frac{x_i}{\mu^2}-c_i+\frac12
    \right]
    \bigg\},
    \label{eq:VCW_second_derivative}
\end{align}
where $a,b\in\{\phi,\varphi\}$ and the NLO rule of Eq.~\eqref{eq:GW_ES_counting} is understood. These equations are the two-derivative special case of the general derivative formulae for the one-loop effective potential~\cite{Camargo-Molina:2016moz}.

The finite counterterm potential is
\begin{align}
    V_{\rm CT}(\phi,\varphi)
    &=
    \frac{\delta m_H^2}{2}\phi^2
    +
    \frac{\delta\lambda_H}{4}\phi^4
    +
    \frac{\delta m_S^2}{2}\varphi^2
    +
    \frac{\delta\lambda_S}{4}\varphi^4
    +
    \frac{\delta\lambda_{HS}}{2}\phi^2\varphi^2
    \nonumber\\
    &\quad
    +
    \delta T_\phi\phi
    +
    \delta T_\varphi\varphi .
    \label{eq:VCT_general}
\end{align}
The linear counterterms are retained explicitly because they impose tadpole conditions without shifting the chosen vacuum coordinates. At $(v,0)$,
\begin{align}
    \left.\partial_\phi V_{\rm CT}\right|_{(v,0)}
    &=
    v\delta m_H^2+v^3\delta\lambda_H+\delta T_\phi,
    &
    \left.\partial_\varphi V_{\rm CT}\right|_{(v,0)}
    &=
    \delta T_\varphi,
    \label{eq:VCT_tadpoles}
    \\
    \left.\partial_\phi^2 V_{\rm CT}\right|_{(v,0)}
    &=
    \delta m_H^2+3v^2\delta\lambda_H,
    &
    \left.\partial_\varphi^2 V_{\rm CT}\right|_{(v,0)}
    &=
    \delta m_S^2+v^2\delta\lambda_{HS},
    \label{eq:VCT_Hessians}
\end{align}
while $\left.\partial_\phi\partial_\varphi V_{\rm CT}\right|_{(v,0)}=0$.

The modified on-shell conditions are imposed on the loop-plus-counterterm part of the potential:
\begin{equation}
    \left.
    \partial_{\chi_i}(V_{\rm CW}+V_{\rm CT})
    \right|_{(v,0)}=0,
    \qquad
    \chi_i\in\{\phi,\varphi\},
    \label{eq:modified_OS_tadpoles}
\end{equation}
\begin{equation}
    \left.
    \partial_{\chi_i}\partial_{\chi_j}(V_{\rm CW}+V_{\rm CT})
    \right|_{(v,0)}
    =
    \begin{pmatrix}
        m_h^2 & 0\\
        0 & 0
    \end{pmatrix}_{ij},
    \label{eq:modified_OS_Hessian}
\end{equation}
where $(\chi_1,\chi_2)=(\phi,\varphi)$. The first entry lifts the flat doublet direction to the observed Higgs-boson mass, while the zero in the singlet entry keeps the tree-level singlet curvature unchanged.

More generally, in a scalar theory with non-trivial tree-level mixing at the
physical vacuum, the on-shell condition preserving the tree-level mixing matrix
would be written in the form
\begin{equation}
    \left.
    \partial_i\partial_j
    \left(
        V_{\rm CW}+V_{\rm CT}
    \right)
    \right|_{\rm vac}
    =
    \left[
        R\,\Delta M_{\rm OS}^2\,R^T
    \right]_{ij},
    \label{eq:general_rotated_OS_condition}
\end{equation}
where \(R\) diagonalises the tree-level scalar mass matrix at the vacuum and
\(\Delta M_{\rm OS}^2\) specifies the loop-generated shifts in the physical
mass basis. This is the form commonly used in multi-scalar models with
non-trivial vacuum mixing. In the present \(\mathbb Z_2\)-preserving branch,
however, the tree-level mass matrix at \((v,0)\) is already diagonal,
Eq.~\eqref{eq:tree_scalar_curvature_vacuum}, so \(R_{\rm vac}=\mathbf 1\)
up to trivial field ordering and signs. Moreover, \(\mathbb Z_2\) evenness
implies
\[
    \left.
    \partial_\varphi\partial_\phi V_{\rm CW}
    \right|_{(v,0)}
    =
    \left.
    \partial_\varphi\partial_\phi V_{\rm CT}
    \right|_{(v,0)}
    =
    0 .
\]
Therefore Eq.~\eqref{eq:modified_OS_Hessian} is the special-case reduction of
the general rotated on-shell condition: it preserves the vanishing physical
Higgs--singlet mixing angle while lifting only the flat Higgs/scalon direction.
The field-dependent mixing angle \(\alpha(\varphi,\phi)\) introduced in
Sec.~\ref{subsec:field-dependent-spectrum} is used to obtain the off-shell
eigenvalues entering \(V_{\rm CW}\), but it is not an additional
renormalisation condition away from the physical vacuum.

We define, all at $(v,0)$,
\begin{equation}
    H_{hh}\equiv \left.\partial_\phi^2V_{\rm CW}\right|,
    \qquad
    H_{ss}\equiv \left.\partial_\varphi^2V_{\rm CW}\right|,
    \qquad
    N_h\equiv \left.\partial_\phi V_{\rm CW}\right|,
    \label{eq:Hhh_Hss_Nh_defs}
\end{equation}
and
\begin{equation}
    H_{GG}\equiv
    \left.
    \frac{\partial^2V_{\rm CW}}{\partial G^0\partial G^0}
    \right|_{(v,0)} .
    \label{eq:HGG_def}
\end{equation}
The Goldstone and Higgs-curvature conditions are
\begin{equation}
    H_{GG}+\delta m_H^2+v^2\delta\lambda_H=0,
    \qquad
    H_{hh}+\delta m_H^2+3v^2\delta\lambda_H=m_h^2.
    \label{eq:Goldstone_Higgs_conditions}
\end{equation}
They fix
\begin{equation}
    \delta\lambda_H
    =
    \frac{m_h^2-H_{hh}+H_{GG}}{2v^2},
    \qquad
    \delta m_H^2
    =
    \frac12\left(H_{hh}-3H_{GG}-m_h^2\right).
    \label{eq:doublet_counterterm_solutions}
\end{equation}
The Higgs-direction tadpole condition gives
\begin{equation}
    \delta T_\phi=vH_{GG}-N_h.
    \label{eq:deltaTphi_solution}
\end{equation}
Thus the Higgs tadpole counterterm is fixed by the renormalisation conditions and should not be set to zero independently. In the present model, however, all NLO field-dependent masses depend on the doublet background only through $H^\dagger H$ (with $m_G^2$ independent of $\phi$ and $F_-\equiv0$ along the electroweak ray), which implies $vH_{GG}=N_h$ and hence $\delta T_\phi=0$ identically; the restriction of the renormalised two-field potential to the ray therefore coincides \emph{exactly} with the one-dimensional construction of Sec.~\ref{subsec:flat-direction-limit}.

Because the one-loop potential is even in $\varphi$, the singlet tadpole and mixed curvature vanish at $(v,0)$, implying
\begin{equation}
    \delta T_\varphi=0.
    \label{eq:deltaTvarphi_solution}
\end{equation}
The singlet-curvature condition fixes only one linear combination,
\begin{equation}
    H_{ss}+\delta m_S^2+v^2\delta\lambda_{HS}=0.
    \label{eq:singlet_curvature_condition}
\end{equation}
We adopt the minimal finite convention
\begin{equation}
    \delta m_S^2=0,
    \qquad
    \delta\lambda_{HS}=-\frac{H_{ss}}{v^2},
    \qquad
    \delta\lambda_S=0.
    \label{eq:minimal_singlet_counterterm_convention}
\end{equation}
The choice $\delta\lambda_S=0$ is a finite convention, since $\lambda_S$ is not fixed by local vacuum conditions at $v_s=0$. Other finite choices, such as $\delta\lambda_{HS}=0$ and $\delta m_S^2=-H_{ss}$, impose the same local renormalisation conditions but define a different off-shell truncated one-loop potential. We therefore regard Eq.~\eqref{eq:minimal_singlet_counterterm_convention} as part of the definition of our scheme.

This point is conceptually important. The modified on-shell prescription restores the measured Higgs curvature by a finite counterterm and therefore differs from the strict Gildener--Weinberg construction, in which $m_h^2=8Bv^2$ would be a loop-level prediction. Our formulation does not predict the Higgs mass from the singlet sector alone. Instead, $v$ and $m_h$ are used as physical renormalisation inputs, while scale invariance forbids an independent singlet mass parameter. Reimposing the same physical conditions under variations of $\mu$ absorbs the explicit scale dependence of the Coleman--Weinberg potential into the finite counterterms at the physical vacuum. The resulting dark-sector predictions are therefore not artefacts of a particular renormalisation-scale choice: their residual scale dependence begins beyond the one-loop accuracy of the calculation. The combined vacuum, scale-invariance and relic-density conditions can consequently be used to identify robust isolated dark-matter solutions. This is the precise sense in which the model is highly constrained without containing an independent dark-matter mass parameter.

Combining the above results, the counterterm potential used below is
\begin{align}
    V_{\rm CT}(\phi,\varphi)
    &=
    \frac14\left(H_{hh}-3H_{GG}-m_h^2\right)\phi^2
    +
    \frac{m_h^2-H_{hh}+H_{GG}}{8v^2}\phi^4
    \nonumber\\
    &\quad
    -
    \frac{H_{ss}}{2v^2}\phi^2\varphi^2
    +
    (vH_{GG}-N_h)\phi .
    \label{eq:VCT_final_minimal}
\end{align}
It ensures that the renormalised potential has a stationary point at $(v,0)$, that the flat doublet direction is lifted to the input Higgs mass, and that the singlet curvature remains equal to $\lambda_{HS}v^2$.

\subsection{Flat-direction limit and analytic vacuum condition}
\label{subsec:flat-direction-limit}

As an analytic check, we restrict Eq.~\eqref{eq:Veff_full_definition} to the electroweak ray $\varphi=0$. At $\mu_\star$, $V_0(\phi,0)=0$, and the one-loop contribution takes the standard Gildener--Weinberg form~\cite{Gildener:1976ih}
\begin{equation}
    V_{\rm CW}^{\rm 1D}(\phi)
    =
    A\phi^4
    +
    B\phi^4\log\frac{\phi^2}{v^2}.
    \label{eq:VCW_1D_AB}
\end{equation}
If $M_i^2(\phi,0)=\kappa_i\phi^2$, then
\begin{equation}
    A=
    \sum_i
    \frac{n_i}{64\pi^2}\kappa_i^2
    \left[
        \log\frac{\kappa_i v^2}{\mu^2}-c_i
    \right],
    \qquad
    B=
    \sum_i
    \frac{n_i}{64\pi^2}\kappa_i^2 .
    \label{eq:AB_coefficients}
\end{equation}
Along the electroweak ray the non-zero coefficients are
\begin{equation}
    \kappa_W=\frac{g^2}{4},
    \qquad
    \kappa_Z=\frac{g^2+g'^2}{4},
    \qquad
    \kappa_t=\frac{y_t^2}{2},
    \qquad
    \kappa_s=\tfrac12\lambda_{HS},
    \label{eq:kappa_values}
\end{equation}
while the Goldstones and the scalon are massless at this order. Therefore,
\begin{equation}
    B
    =
    \frac{1}{64\pi^2v^4}
    \left(
        6m_W^4+3m_Z^4-12m_t^4+m_S^4
    \right),
    \qquad
    m_S^2=\tfrac12\lambda_{HS}v^2.
    \label{eq:B_explicit_masses}
\end{equation}
The positive singlet contribution must compensate the negative top-quark term for the flat direction to be lifted upward.

The corresponding one-dimensional effective potential may be written as
\begin{equation}
    V_{\rm eff}^{\rm 1D}(\phi)
    =
    A\phi^4+B\phi^4\log\frac{\phi^2}{v^2}
    +
    \frac{\delta m_H^2}{2}\phi^2
    +
    \frac{\delta\lambda_H}{4}\phi^4,
    \label{eq:Veff_1D}
\end{equation}
 where no linear counterterm appears because $\delta T_\phi=0$ identically in this model (cf.\ Sec.~\ref{subsec:modified-OS-counterterms}), so that Eq.~\eqref{eq:Veff_1D} is the exact restriction of the two-field scheme to the electroweak ray. Imposing
\begin{equation}
    \left.\frac{dV_{\rm eff}^{\rm 1D}}{d\phi}\right|_{\phi=v}=0,
    \qquad
    \left.\frac{d^2V_{\rm eff}^{\rm 1D}}{d\phi^2}\right|_{\phi=v}=m_h^2
    \label{eq:1D_OS_conditions}
\end{equation}
gives
\begin{equation}
    \frac{\delta m_H^2}{2}
    =
    \frac14(8Bv^2-m_h^2),
    \qquad
    \frac{\delta\lambda_H}{4}
    =
    -A-\frac32B+\frac{m_h^2}{8v^2}.
    \label{eq:1D_counterterms}
\end{equation}
Substitution into Eq.~\eqref{eq:Veff_1D} yields
\begin{equation}
    V_{\rm eff}^{\rm 1D}(v)
    =
    \frac12Bv^4-\frac18m_h^2v^2,
    \qquad
    V_{\rm eff}^{\rm 1D}(0)=0.
    \label{eq:Veff_v_1D}
\end{equation}
With the mass normalisation $m_S^2=\lambda_{HS}v^2/2$, Eq.~\eqref{eq:Veff_v_1D} gives
\begin{equation}
    V_{\rm eff}(v,0)=V_{\rm eff}(0,0)
    \quad\Longrightarrow\quad
    m_S\simeq632~{\rm GeV},
    \label{eq:vacuum_degeneracy_632}
\end{equation}
for the numerical inputs used in the LanHEP/CalcHEP implementation. At this point $\lambda_{HS}=2m_S^2/v^2\simeq13.2$, and the same diagnostic gives $\lambda_S^{\rm eff}\simeq24$. 
Thus the electroweak minimum is deeper than the trivial vacuum at $(\phi,\varphi)=(0,0)$ along the flat direction if
\begin{equation}
    m_h^2>4Bv^2.
    \label{eq:analytic_vacuum_depth_condition}
\end{equation}
This is only a one-dimensional check. The full analysis must still exclude deeper stationary points with $\varphi\neq0$ in the two-field potential. 

In the conventional Gildener--Weinberg treatment without imposing the Higgs mass through a finite curvature counterterm, the scalon mass is predicted as
\begin{equation}
    m_{\rm scalon}^2=8Bv^2.
    \label{eq:GW_scalon_mass}
\end{equation}
In the present modified on-shell formulation, $m_h$ is instead used as an input renormalisation condition, so Eq.~\eqref{eq:GW_scalon_mass} is not imposed as a prediction.

\subsection{LanHEP/CalcHEP implementation}
\label{app:lanhep-calchep}

This subsection summarises the LanHEP/CalcHEP implementation used for the dark-matter calculation. The implementation follows the scalar-potential normalisation
\begin{equation}
    V_0(H,S)=
    \lambda_H(H^\dagger H)^2
    +\frac{\lambda_{HS}}{2}(H^\dagger H)S^2
    +\frac{\lambda_S}{4}S^4 .
    \label{eq:lanhep_scalar_potential}
\end{equation}
After electroweak symmetry breaking,
\begin{equation}
    H^\dagger H=\frac{(v+h)^2}{2},
\end{equation}
the singlet mass is
\begin{equation}
    m_S^2=\frac{1}{2}\lambda_{HS}v^2 .
    \label{eq:lanhep_mass_relation}
\end{equation}
The LanHEP input
\begin{verbatim}
parameter lhs=2*MX**2/(vevh)**2.
\end{verbatim}
therefore implements
\begin{equation}
    \lambda_{HS}=\frac{2M_X^2}{v^2},
    \qquad
    m_S^2=M_X^2 .
    \label{eq:lanhep_lhs_relation}
\end{equation}
Thus the mass parameter passed to CalcHEP and micrOMEGAs is the singlet mass used in the phenomenological analysis.

The renormalised one-loop potential used in the analysis is
\begin{equation}
    V_{\rm eff}(\phi,\varphi)
    =
    V_0(\phi,\varphi)
    +
    V_{\rm CW}^{\rm NLO}(\phi,\varphi)
    +
    V_{\rm CT}(\phi,\varphi),
    \label{eq:lanhep_veff_definition}
\end{equation}
where $\phi=v+h$ is the neutral Higgs background and $\varphi=S$ is the singlet background. The effective Higgs-portal coupling is defined from the renormalised $hSS$ vertex,
\begin{equation}
    \lambda_{HS}^{\rm eff}
    =
    \frac{1}{v}
    \left.
    \frac{\partial^3 V_{\rm eff}}
    {\partial h\,\partial S\,\partial S}
    \right|_{h=0,\,S=0}.
    \label{eq:lanhep_lhseff_definition}
\end{equation}
Equivalently,
\begin{equation}
    {\cal L}\supset
    -\frac{1}{2}\lambda_{HS}^{\rm eff}v hS^2 .
    \label{eq:lanhep_hss_vertex}
\end{equation}
This coupling enters the freeze-in decay width $h\to SS$ and the Higgs-mediated direct-detection amplitude. Beyond tree level, $\lambda_{HS}^{\rm eff}$ is not fixed by the mass relation in Eq.~\eqref{eq:lanhep_mass_relation}; the mass is fixed by $\lambda_{HS}$, while the interaction vertex is fixed by $\lambda_{HS}^{\rm eff}$.

In the LanHEP model the singlet field is denoted by \texttt{'\textasciitilde X'}, and the Higgs-doublet contraction $H^\dagger H$ is denoted by \texttt{shd*shD}. The portal interaction is implemented as
\begin{verbatim}
lterm -(lhseff/2)*(shd*shD-(lhseff-lhs)/lhseff*vevh**2/2)*'~X'**2.
\end{verbatim}
In field notation this is
\begin{equation}
    {\cal L}_{HS}
    =
    -\frac{\lambda_{HS}^{\rm eff}}{2}
    \left[
    H^\dagger H
    -
    \frac{\lambda_{HS}^{\rm eff}-\lambda_{HS}}
         {\lambda_{HS}^{\rm eff}}
    \frac{v^2}{2}
    \right]S^2 .
    \label{eq:lanhep_shifted_portal}
\end{equation}
The first term in the bracket gives the usual Higgs-portal interaction with the effective coupling $\lambda_{HS}^{\rm eff}$. The second term is a constant subtraction. Its role is to keep the singlet mass fixed by $\lambda_{HS}$, while allowing the $hSS$ vertex to be controlled by $\lambda_{HS}^{\rm eff}$.

At the electroweak vacuum, $H^\dagger H=v^2/2$, and the bracket in Eq.~\eqref{eq:lanhep_shifted_portal} becomes
\begin{equation}
    \frac{v^2}{2}
    -
    \frac{\lambda_{HS}^{\rm eff}-\lambda_{HS}}
         {\lambda_{HS}^{\rm eff}}
    \frac{v^2}{2}
    =
    \frac{\lambda_{HS}}{\lambda_{HS}^{\rm eff}}
    \frac{v^2}{2}.
\end{equation}
The mass term is therefore
\begin{equation}
    {\cal L}_{\rm mass}
    =
    -\frac{\lambda_{HS}^{\rm eff}}{2}
    \frac{\lambda_{HS}}{\lambda_{HS}^{\rm eff}}
    \frac{v^2}{2}S^2
    =
    -\frac{1}{2}M_X^2S^2,
    \label{eq:lanhep_mass_term}
\end{equation}
whereas expanding $H^\dagger H=(v+h)^2/2$ gives the trilinear interaction in Eq.~\eqref{eq:lanhep_hss_vertex}. Thus the shifted portal term reproduces both the required singlet mass and the required effective $hSS$ vertex.

The effective portal coupling used in the scan is related to the input quartics by
\begin{equation}
    \lambda_{HS}^{\rm eff}
    =
    \lambda_{HS}
    +
    \frac{3\lambda_{HS}}{16\pi^2}
    \left(2\lambda_{HS}+3\lambda_S\right).
    \label{eq:lanhep_lhseff_relation}
\end{equation}
Solving this relation for $\lambda_S$ gives
\begin{equation}
    \lambda_S
    =
    \frac{1}{3}
    \left[
    \frac{16\pi^2\left(\lambda_{HS}^{\rm eff}-\lambda_{HS}\right)}
         {3\lambda_{HS}}
    -2\lambda_{HS}
    \right],
    \label{eq:lanhep_ls_solution}
\end{equation}
which is implemented as
\begin{verbatim}
parameter ls=((lhseff-lhs)*(16*PI**2)/(3*lhs) - 2*lhs)/3.
\end{verbatim}

The singlet self-coupling follows the convention
\begin{equation}
    V_{\rm eff}\supset
    \frac{\lambda_S^{\rm eff}}{4}S^4 .
    \label{eq:lanhep_lseff_potential}
\end{equation}
Therefore
\begin{equation}
    \lambda_S^{\rm eff}
    =
    \frac{1}{6}
    \left.
    \frac{\partial^4 V_{\rm eff}}
    {\partial S\,\partial S\,\partial S\,\partial S}
    \right|_{h=0,\,S=0},
    \label{eq:lanhep_lseff_definition}
\end{equation}
since
\begin{equation}
    \frac{\partial^4}{\partial S^4}
    \left(
    \frac{\lambda_S^{\rm eff}}{4}S^4
    \right)
    =
    6\lambda_S^{\rm eff}.
\end{equation}
The one-loop expression used for this quartic coefficient is
\begin{equation}
    \lambda_S^{\rm eff}
    =
    \lambda_S
    +
    \frac{3\lambda_{HS}\left(\lambda_{HS}+2\lambda_S\right)}{8\pi^2}
    +
    \frac{
    \left(-2\lambda_{HS}^2+9\lambda_S^2\right)
    \log\left(\lambda_{HS}/2\right)}
    {16\pi^2}.
    \label{eq:lanhep_lseff_relation}
\end{equation}
This is implemented as
\begin{verbatim}
parameter lseff=ls + 3*lhs*(lhs+2*ls)/(8*PI**2) 
                   + ((-2*lhs**2+9*ls**2)*log(lhs/2))/(16*PI**2).
\end{verbatim}
Together with
\begin{verbatim}
lterm -(lseff/4)*'~X'**4.
\end{verbatim}
this gives Eq.~\eqref{eq:lanhep_lseff_potential}.

The one-dimensional vacuum-energy diagnostic used to check the upper bound can be implemented in LanHEP as
\begin{verbatim}
parameter Bveff  = (6*MW**4+3*MZ**4-12*Mtp**4+MX**4)/(64*PI**2*vevh**4):
          '1D CW coefficient along S=0',
          
          VeffEW = Bveff*vevh**4/2-Mh**2*vevh**2/8:
          'Veff(v,0) with Veff(0,0)=0',

          MXdeg  = sqrt(sqrt(16*PI**2*Mh**2*vevh**2-6*MW**4-3*MZ**4+12*Mtp**4)):
          'mass where VeffEW=0',

          lhsdeg = 2*MXdeg**2/vevh**2:
          'lambda_HS at VeffEW=0',

          lsdeg  = ((lhseff-lhsdeg)*(16*PI**2)/(3*lhsdeg)-2*lhsdeg)/3:
          'lambda_S at VeffEW=0',

          lseffdeg = lsdeg+3*lhsdeg*(lhsdeg+2*lsdeg)/(8*PI**2)
                     +((-2*lhsdeg**2+9*lsdeg**2)*log(lhsdeg/2))/(16*PI**2):
          'lambda_S_eff at VeffEW=0'.
\end{verbatim}
For the numerical inputs of the CalcHEP model this gives $M_X\simeq632.2$~GeV, $\lambda_{HS}\simeq13.2$ and $\lambda_S^{\rm eff}\simeq24$. This diagnostic checks the electroweak-ray condition $V_{\rm eff}(v,0)=V_{\rm eff}(0,0)$.

The LanHEP/CalcHEP implementation used in this work is publicly available in the High Energy Physics Model DataBase (HEPMDB) at
\url{https://hepmdb.soton.ac.uk/hepmdb:0826.0358}.

\newpage
\bibliography{references}

\end{document}